# Superconductivity in a Layered Ta$_4$Pd$_3$Te$_{16}$ with PdTe$_2$ Chains

Wen-He Jiao,[†] Zhang-Tu Tang,[†] Yun-Lei Sun,[†] Yi Liu,[†] Qian Tao,[†] Chun-Mu Feng,[†] Yue-Wu Zeng,[#] Zhu-An Xu,[†,‡] and Guang-Han Cao[*,†,‡]

[†] Department of Physics, [‡] State Key Lab of Silicon Materials and [#] Center of Electron Microscope, Zhejiang University, Hangzhou 310027, China

Supporting Information Placeholder

**ABSTRACT:** Superconductivity in low-dimensional compounds has long attracted much interest. Here we report superconductivity in a low-dimensional ternary telluride Ta$_4$Pd$_3$Te$_{16}$ in which the repeating layers contain edge-sharing octahedrally-coordinated PdTe$_2$ chains along the crystallographic $b$ axis. Measurements of electrical resistivity, magnetic susceptibility and specific heat on the Ta$_4$Pd$_3$Te$_{16}$ crystals, grown via a self-flux method, consistently demonstrate bulk superconductivity at 4.6 K. Further analyses of the data indicate significant electron-electron interaction, which allows electronic Cooper pairing in the present system.

Recent decades witness increasing numbers of brand-new superconductors including the two classes of high-temperature superconductors ever discovered[1,2] in low-dimensional systems. Generally speaking, low dimensionality leads to special electronic structures and allows relatively strong fluctuations, which may enhance superconductivity, although charge-density wave (CDW) sometime competes, especially in the quasi-one-dimensional case.[3] Thus it is rational to explore new superconductors in materials with low-dimensional crystal structures. From a crystal-chemistry viewpoint, low-dimensional structures easily appear in an inorganic metal compound whose constituent anions have relatively low electronegativity and high polarizability. For example, metal chalcogenides, as compared with metal oxides, tend to crystallize in layered structures,[4] which give rise to rich interesting physical phenomena including CDW and superconductivity.[3,5] Among chalcogenides, tellurides often differ from sulfides and selenides in crystal structures, electronic structures and physical properties, primarily because of the diffuse nature of the tellurium valence orbitals.

While binary chalcogenides such as NbSe$_2$[6] and NbSe$_3$[7] were intensively studied because of their CDW and superconductivity, ternary and multi-element chalcogenides have not been paid much attention since their births in the 1980s and 1990s.[4,8-12] Very recently, superconductivity was discovered in Pd-based ternary chalcogenides $M_2$Pd$_x$Q$_5$ ($M$=Nb and Ta, $Q$=S and Se)[13-17] and Nb$_3$Pd$_x$Se$_7$.[18] The crystal structure consists of corrugated $M$-Pd-$Q$ layers in which the $M$-$Q$ chains are interconnected abreast by *square-planarly* coordinated Pd atoms.[8] The most remarkable property of the new superconductors is an extremely high upper critical field along the chain direction. Furthermore, the material was notably found to have potential industrial applications by the successful fabrication of flexible Nb$_2$Pd$_x$S$_{5-\delta}$ superconducting fibers.[17] Nevertheless, none of their analogous tellurides has been synthesized because of the distinct feature of tellurium mentioned above.

Here we report superconductivity in a ternary Pd-based *telluride* Ta$_4$Pd$_3$Te$_{16}$ which was first synthesized and characterized by Mar and Ibers in 1991.[9] The compound crystallizes in space group $I2/m$ with a monoclinic unit cell of $a$=17.687(4) Å, $b$=3.735(1) Å, $c$=19.510(4) Å, and $\beta$=110.42° at 111 K. Compared to the aforementioned $M_2$Pd$_x$Q$_5$ phase, the crystal structure has relatively flat Ta-Pd-Te layers, as illustrated in Figure 1a. The most remarkable difference, however, is that the Pd atoms are *octahedrally* coordinated, forming edge-sharing PdTe$_2$ chains along the $b$ axis. This difference gives rise to totally different crystal field for Pd atoms, which possibly brings about a disparate electronic state. The PdTe$_2$ chains are separated by TaTe$_3$ chains (with face-sharing bicapped trigonal prisms) and Ta$_2$Te$_4$ double chains (with edge-sharing distorted octahedra). Therefore, Ta$_4$Pd$_3$Te$_{16}$ is a layered compound with quasi-one-dimensional characteristics.

The Ta$_4$Pd$_3$Te$_{16}$ crystals were grown by a self-flux method, similar to previous report,[9] but the growth parameters were varied. Powders of the elements Ta (99.97%), Pd (99.995%) and Te (99.99%), all from Alfa Aesar, in an atomic ratio of Ta:Pd:Te = 2:3:15 were thoroughly mixed together, then loaded, and sealed into a quartz ampoule evacuated. The ampoule was slowly heated up to 1223 K, holding for 24 h. After that, it was allowed to cool to 923 K at a rate of 5 K/h, followed by cooling down to room temperature. Shinning gray-black soft crystals in flattened needle shape were harvested with a typical dimension of 2.5×0.25×0.1 mm$^3$, as shown in the left panel of Figure 1b. The scanning-electron-microscope (SEM)[19] image shown in the chain direction. The chemical composition was determined by energy dispersive X-ray spectroscopy (EDXS),[19] which gives the atomic ratio Ta:Pd:Te=4:3:16 within the measurement precision (±3-5% depending on the elements measured). This result indicates that there is no significant deficiency for Pd, contrasting with the cases of the $M_2Pd_xQ_5$ and $Nb_3Pd_xSe_7$ phases mentioned above. Shown in the left panel of Figure 1c is the X-ray diffraction (XRD)[19] pattern at 298 K by a conventional θ-2θ scan for the crystals lying on a sample holder. Only multiple reflections of (-103) planes can be seen, consistent with the layered crystal structure depicted in Figure 1a. The inter-plane spacing is determined to be 6.530 Å, close to the calculated value of 6.502 Å using the reported cell parameters at 111 K,[9] pointing to the $Ta_4Pd_3Te_{16}$ phase. The full width at half maximum (FWHM) is only 0.04°, e.g., for the (-309) peak, indicating high quality of the crystals. We also employed high-resolution transmission electron microscopy (HRTEM)[19] to confirm the structure further. As shown in the right panel of Figure 1c, the lattice parameter $b$ measured in the [-103] zone image is ~3.76(5) Å, close to the reported value as well. The HRTEM image is found to be completely consistent with the crystal structure of $Ta_4Pd_3Te_{16}$, as illustrated in the upper right inset.

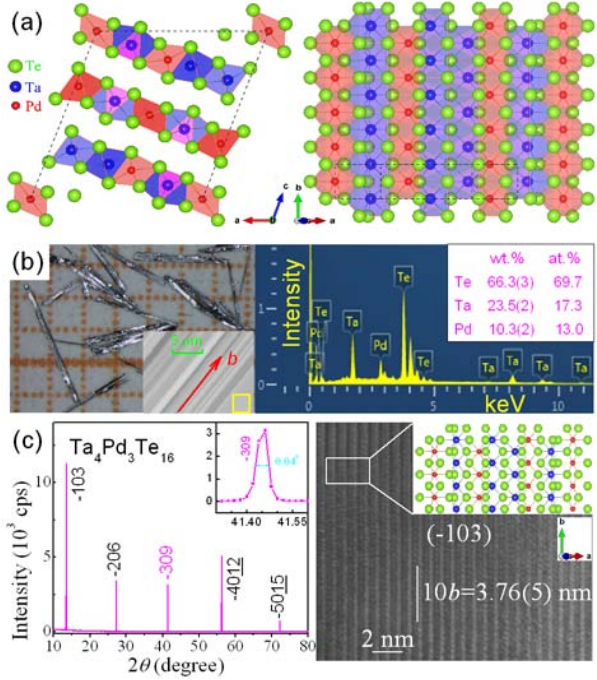

**Figure 1.** (a) Crystal structure of $Ta_4Pd_3Te_{16}$ projected along the [010] (left) and [-103] (right) directions. (b) Photographs of the as-grown $Ta_4Pd_3Te_{16}$ crystals on a mm-grid paper (left panel) and a typical image under a scanning electron microscope (inset of the panel), on which energy dispersive x-ray spectra (right panel) were collected. (c) X-ray diffraction pattern (left) and image of (-103) planes (right) with a high-resolution transmission electron microscope at room temperature for the $Ta_4Pd_3Te_{16}$ crystalline flakes lying on sample holders. The inset of the left panel zooms in the (-309) reflection, and the inset of the right panel expands the real structure corresponding to the area marked.

Figure 2a-2c show temperature ($T$) dependence of electrical resistivity along the $b$ axis ($\rho_b$)[20] for the $Ta_4Pd_3Te_{16}$ crystals. The room-temperature $\rho_b$ is $6.1\times10^{-7}$ Ω·m, comparable to that of a normal metal. With decreasing $T$, $\rho_b(T)$ decreases almost linearly down to ~50 K without any sharp anomaly due to a possible CDW transition. The $\rho_b(T)$ data from 6 K to 35 K obeys $\rho_b=\rho_0+AT^\alpha$ with $\alpha=2.00$ (Figure 2c), which is one of the hallmarks of Fermi liquid, suggesting dominant electron-electron scattering at low temperatures. The residual resistivity $\rho_0$ by the extrapolation turns out to be $2.34\times10^{-8}$ Ω·m. Consequently, the residual resistivity ratio (RRR, defined as the ratio of the resistivity at room temperature and at zero temperature) is 26. It is much larger than those (~2) of the $M_2Pd_xQ_5$ compounds,[14,15,17] in which the Pd deficiency may cause impurity scatterings. The much less impurity scattering in our $Ta_4Pd_3Te_{16}$ crystals meets with the exact stoichiometry without Pd deficiency evidenced from the above EDXS result.

A superconducting transition takes place below 5 K, as clearly shown in Figure 2b. The onset, midpoint, and zero-resistance temperatures are 4.62 K, 4.46 K and 4.11 K, respectively. Superconductivity is confirmed by dc magnetic susceptibility measurement[20] (Figure 2d), showing strong diamagnetism below 4.6 K. Since the zero-field-cooling (ZFC) signal is a measure of magnetic shielding and, the sample was a crystal, the superconducting volume fraction can be estimated to be 90%±4% (the error mainly comes from the weighing accuracy). The field-cooling (FC) diamagnetic signal is obviously weaker due to flux pinning effect. Nevertheless, it is still much stronger than that in $M_2Pd_xQ_5$ system,[14,15,17] indicating comparatively weak flux pinning in $Ta_4Pd_3Te_{16}$. Other superconducting properties such as critical magnetic fields will be reported elsewhere.[21]

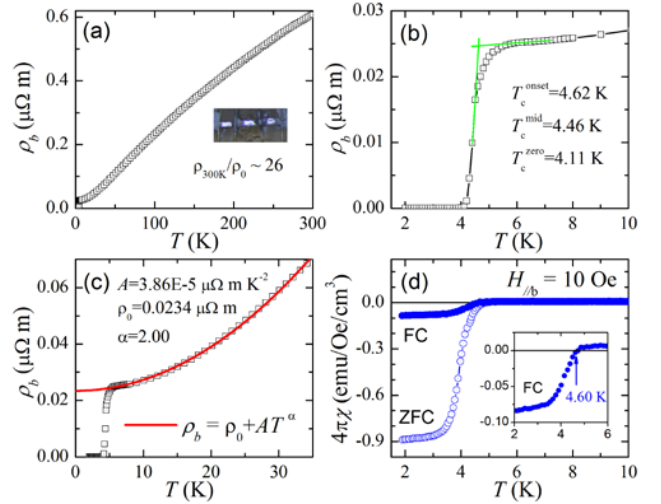

**Figure 2.** (a-c) Temperature dependence of electrical resistivity along the $b$ axis for $Ta_4Pd_3Te_{16}$ crystals. (d) Superconducting transition in dc magnetic susceptibility. Both zero-field cooling (ZFC) and field cooling (FC, shown more clearly in the inset) curves were measured.

Low-$T$ specific heat, $C(T)$, may supply important information to a superconductor. At temperatures much below Debye temperature $\Theta_D$, the specific heat contributed from phonons can be expressed by Debye $T^3$ law: $C_{ph}=\beta T^3$. Thus the total specific heat measured[20] can be expressed by $C=\gamma T+\beta T^3$, and the plot of $C/T$ vs. $T^2$ allows a linear fitting of normal-state $C(T)$ of the Ta$_4$Pd$_3$Te$_{16}$ crystals, as shown in Figure 3a. The fitting yields a Sommerfeld parameter γ=42.8±1.6 mJ·K$^{-2}$·mol$^{-1}$ and a phononic coefficient β=11.11±0.05 mJ·K$^{-4}$·mol$^{-1}$. The Θ$_D$ value is then determined to be 159.1±0.3 K by the formula Θ$_D$=[(12/5)NRπ$^4$/β]$^{1/3}$, where N counts the number of atoms per formula unit (fu), and R is the gas constant. The density of states (DOS) at Fermi level, N(E$_F$), is also estimated to be 18.1±0.7 eV$^{-1}$·fu$^{-1}$ by the relation N(E$_F$)=3γ/(k$_B^2$π$^2$) for the case of non-interacting electron systems.

The electronic specific heat in the superconducting state can be obtained by a simple subtraction: C$_{el}$=C−βT$^3$. Figure 3b plots C$_{el}$/T as a function of temperature. A clear specific-heat jump, ΔC$_{el}$, shows up below 4.6 K, which further confirms bulk superconductivity. The ΔC$_{el}$/T$_c$ value is estimated to be 54±3 mJ·K$^{-2}$·mol$^{-1}$ and, by an entropy conserving construction, the midpoint temperature of the thermodynamic transition is determined to be 4.1 K, coincident with the zero-resistance temperature. The C$_{el}$(T) data below 1 K tend to saturate, suggesting that fully gapped superconductivity is likely. However, with the limited data of ultra-low temperatures, one cannot rule out the possibility of nodal and/or multiple superconducting gaps owing to the complexity of the Fermi surfaces.[22] In the case of full-gap scenario, the residual electronic specific-heat coefficient is 7(1) mJ·K$^{-2}$·mol$^{-1}$ at T→0 K, indicating presence of non-superconducting metallic portion in the sample. Assuming that the non-superconducting phase has the same γ-value, the superconducting molar fraction is then estimated to be 84%±6%, basically consistent with the above magnetic measurement result.

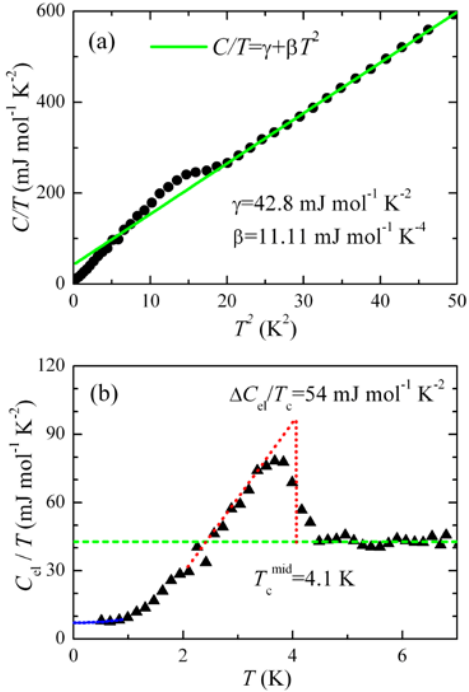

**Figure 3.** Temperature dependence of specific heat for Ta$_4$Pd$_3$Te$_{16}$ crystals. (a) C/T vs. T$^2$, in which the green straight line is the fit with C/T=γ+βT$^2$ for the normal-state data from 4.5 K to 7 K. (b) Plot of C$_{el}$/T vs. T, where C$_{el}$=C−βT$^3$.

Taken the superconducting fraction of 90% into consideration for correction, the dimensionless specific-heat jump ΔC/(γT$_c$) is 1.4±0.2, which agrees well with the BCS theoretical value of 1.43, suggesting a weak coupling scenario. On the other hand, the electron-phonon (e-ph) coupling constant λ$_{ph}$ can be estimated by McMillan formula,[23]

$$\lambda_{ph} = \frac{1.04 + \mu^* \ln(\Theta_D / 1.45T_c)}{(1 - 0.62\mu^*)\ln(\Theta_D / 1.45T_c) - 1.04},$$

where the Coulomb repulsion parameter μ* is set to be 0.13 empirically. This resultant λ$_{ph}$ is 0.77, suggesting that Ta$_4$Pd$_3$Te$_{16}$ belongs to an intermediately coupled superconductor. In addition, the e-ph coupling strength can also be evaluated by comparing the DOS from specific-heat measurement and the "bare" DOS from band-structure calculations, N$_{bs}$(E$_F$). Using the calculation result of N$_{bs}$(E$_F$)=5.5 eV$^{-1}$·fu$^{-1}$,[22] and applying the relation N(E$_F$)/N$_{bs}$(E$_F$)=1+λ, one obtains an unusually large coupling constant of λ=2.3. This points to electron-nonphonon couplings, λ$_{nph}$=λ−λ$_{ph}$=1.53, which is about twice of λ$_{ph}$ itself, in accordance with the dominant electron-electron scattering revealed by the T$^2$-dependence in low-T resistivity.

The significant electron-electron interactions are also manifested by the criteria of Kadowaki-Woods ratio,[24] defined as A/γ$^2$. Note that there are three Pd atoms in one chemical formula of Ta$_4$Pd$_3$Te$_{16}$, hence the Sommerfeld parameter should be reasonably considered as γ$_{Pd}$=14.3±0.5 mJ·K$^{-2}$·mol-Pd$^{-1}$, which is close to those of M$_2$Pd$_x$Q$_5$[13,15,16] and cobalt oxyhydrate[25] superconductors. In this circumstance, the A/γ$_{Pd}^2$ value is calculated to be 19 μΩ·cm·mol$^2$·K$^2$·J$^{-2}$, placing this material into the group of strongly correlated electron systems including heavy fermion compounds.[24]

Superconductivity in the quasi-one-dimensional Ta$_4$Pd$_3$Te$_{16}$ reminds us of possible excitonic superconducting mechanism, first proposed by Little in 1964.[26] Such an unconventional electronic Cooper pairing by exchanging excitons was hotly discussed recently,[27] and the stabilization of excitons was probed[28] in Ta$_2$NiSe$_5$ which has a similar quasi-one-dimensional crystal structure.[10] In Ta$_4$Pd$_3$Te$_{16}$, charge transfer from the Te$^{2-}$ to Pd$^{4+}$/Ta$^{5+}$ was evident according to the band-structure calculations.[22] Taken the remarkable electron-electron interaction into consideration together, the unconventional electronic Cooper pairing is then possible, which deserves further investigations.

To our knowledge, only few multi-element telluride (except for doped or intercalation binary tellurides) has been discovered to be a superconductor so far.[29] On the other hand, dozens of layered multi-element tellurides have been synthesized,[4,11-12] but they have not been well studied either by chemical doping/intercalation or by physical property measurements. We hope that the discovery of superconductivity in Ta$_4$Pd$_3$Te$_{16}$ will stimulate exploration of superconductivity for the future in numerous multi-element layered tellurides.

# AUTHOR INFORMATION

## Corresponding Author

* E-mail: ghcao@zju.edu.cn

## ACKNOWLEDGMENT

We acknowledge the supports from the National Basic Research Program of China (Grant Nos. 2011CBA00103 and 2010CB923003), the NSF of China (Contract No. 11190023), and the Fundamental Research Funds for the Central Universities of China.

## REFERENCES

(1) Bednorz, J. G.; Müller, K. A. *Z. Phys. B* **1986**, *64*, 189.
(2) Kamihara, Y.; Watanabe, T.; Hirano, M.; Hosono, H. *J. Am. Chem. Soc.* **2008**, *130*, 3296.
(3) For a recent review, see: Smaalen, S. V. *Acta. Cryst.* **2005**, *A61*, 51.
(4) (a) Pell, M. A.; Ibers, J. A. *Chem. Ber./Recueil* **1997**, *130*, 1. (b) Mitchell, K.; Ibers, J. A. *Chem. Rev.* **2002**, *102*, 102.
(5) Wilson, J. A.; Disalvo, F. J.; Mahajan, S. *Adv. Phys.* **1975**, *24*, 117.
(6) (a) Matthias, B. T.; Geballe, T. H.; Campton, V. B. *Rev. Mod. Phys.* **1963**, *35*, 1. (b) Malliakas, C. D.; Kanatzidis, M. G. *J. Am. Chem. Soc.* **2013**, *135*, 1710.
(7) Monceau, P.; Ong, N. P.; Portis, A. M.; Meerschaut, A.; Rouxel, J. *Phys. Rev. Lett.* **1976**, *37*, 602.
(8) (a) Keszler, D. A.; Ibers, J. A.; Shang, M.; Lu, J. *J. Solid State Chem.* **1985**, *57*, 68. (b) Squattritto, P. J.; Sunshine, S. A.; Ibers, J. A. *J. Solid State Chem.* **1986**, *64*, 261.
(9) Mar, A.; Ibers, J. A. *J. Chem. Soc., Dalton Trans.* **1991**, 639.
(10) Sunshine, S. A.; Ibers, J. A. *Inorg. Chem.* **1985**, *24*, 3611.
(11) (a) Mar, A.; Jobic, S.; Ibers, J. A. *J. Am. Chem. Soc.* **1992**, *114*, 8963. (b) Mar, A.; Ibers, J. A. *J. Am. Chem. Soc.* **1993**, *115*, 3227. (c) Pell, M. A.; Ibers, J. A. *J. Am. Chem. Soc.* **1995**, *117*, 6284. (d) Liimatta, E. W.; Ibers, J. A. *J. Solid State Chem*. 1989, *78*, 7. 4316.
(12) (a) Tremel, W. *Angew. Chem., Int. Ed. Engl.* **1991**, *30*, 840. (b) Tremel, W. *Angew. Chem., Int. Ed. Engl.* **1993**, *32*, 1752. (c) Tremel, W. *J. Chem. Soc., Chem. Commun.* **1991**, 1405.
(13) Zhang, Q.; Li, G.; Rhodes, D.; Kiswandhi, A.; Besara, T.; Zeng, B.; Sun, J.; Siegrist, T.; Johannes, M. D.; Balicas, L. *Sci. Rep.* **2013**, *3*, 1446.
(14) Lu, Y. F.; Takayama, T.; Bangura, A. F.; Katsura, Y.; Hashizeme, D.; Takagi, H. arXiv 1308.3766.
(15) Kim, S.; Lee, B.; Choi, K. Y.; Jeon, B. G.; Jang, D. H.; Patil, D.; Patil, S.; Kim, R.; Choi, E. S.; Lee, S.; Yu, J.; Kim, K. H. arXiv 1310.5975.
(16) Niu, C. Q.; Yang, J. H.; Li, Y. K.; Chen, B.; Zhou, N.; Chen, J.; Jiang, L. L.; Chen, B.; Yang, X. X.; Cao, C.; Dai, J. H.; Xu, X. F. *Phys. Rev. B* **2013**, *88*, 104507.
(17) Yu, H.; Zuo, M.; Zhang, L.; Tan, S.; Zhang, C.; Zhang, Y. *J. Am. Chem. Soc.* **2013**, *135*, 12987.
(18) Zhang, Q.; Rhodes, D.; Zeng, B.; Besara, T.; Siegrist, T.; Johannes, M. D.; Balicas, L. *Phys. Rev. B* **2013**, *88*, 024508.
(19) The SEM observation and the EDXS measurement were carried out on a field emission scanning electron microscope (FEI Model SIRION). The XRD experiment was performed at room temperature with a monochromatic Cu K$_{\alpha 1}$ radiation using a PANalytical x-ray diffractometer (Model EMPYREAN). We employed a conventional $\theta$-$2\theta$ scan for the as-grown crystalline flakes lying on a sample holder. HRTEM observation was conducted with conventional experimental procedures employing a FEI Tecnai G$^2$ F20 scanning transmission electron microscope.
(20) The electrical resistivity was measured down to 1.8 K on a Quantum Design Physical Property Measurement System (PPMS-9) by a standard four-terminal method. The four electrodes attaching to gold wires with silver paste were made linearly along the needle direction (see the inset of Figure 2a). Thus we actually measured the electrical resistivity along the crystallographic *b* axis. The size of the electrical contact pads leads to a total uncertainty in the absolute values of resistivity of $\pm 30\%$. The dc magnetic susceptibility was measured on a Quantum Design Magnetic Property Measurement System (MPMS-5). A selected needle-like crystal (0.23 mg) was carefully installed on a sample holder with the applied field parallel to the needle direction. In this case, demagnetization effect is negligible. Both the zero-field-cooling (ZFC) and field-cooling (FC) protocols were employed under a field of 10 Oe. The heat capacity of a dozen of crystals (1.58 mg) placed side by side on the sample holder was measured by a relaxation method down to 0.5 K on the PPMS-9 machine.
(21) Our preliminary result shows that Ta$_4$Pd$_3$Te$_{16}$ is an extreme type-II ($H_{c2} \gg H_{c1}$) superconductor with remarkable anisotropy in upper critical field.
(22) Alemany, P.; Jobic, S.; Brec, R.; Canadell, E. *Inorg. Chem.* **1997**, *36*, 5050.
(23) McMillan, W. L. *Phys. Rev.* **1968**, *167*, 331.
(24) (a) Rice, M. J. *Phys. Rev. Lett.* **1968**, *20*, 1439. (b) Kadowaki, K.; Woods, S. B. *Solid State Commun.* **1986**, *58*, 507.
(25) Cao, G.; Feng, C.; Xu, Y.; Lu, W.; Shen, J.; Fang, M.; Xu, Z. *J. Phys.: Condens. Matter* **2003**, *15*, L519.
(26) Little, W. A. *Phys. Rev.* **1964**, *134*, A1416.
(27) (a) Wezel, J. v.; Nahai-Williamson, P.; Saxena, S. *Phys. Rev. B.* **2011**, *83*, 024502. (b) Seki, K.; Eder, R.; Ohta, Y. *Phys. Rev. B.* **2011**, *84*, 245106. (c) Zenker, B.; Ihle, D.; Bronold, F. X.; Fehske, H. *Phys. Rev. B.* **2012**, *85*, 121102(R).
(28) Wakisaka, Y.; Sudayama, T.; Takubo, K.; Mizokaw, T.; Arita, M.; Namatame, H.; Taniguchi, M.; Katayama, N.; Nohara, M.; Takagi, H. *Phys. Rev. Lett.* **2009**, *103*, 026402.
(29) Superconductivity was recently discovered in a ternary telluride CsBi$_4$Te$_6$ as a low-dimensional narrow-gap semiconductor. See: Malliakas, C. D.; Chung, D. Y.; Claus, H.; Kanatzidis, M. G. *J. Am. Chem. Soc.* **2013**, *135*, 14540.